**Near-Infrared Fluorescence Enhanced (NIR-FE) Molecular Imaging of Live Cells on Gold Substrates****

*Guosong Hong, Scott M. Tabakman, Kevin Welsher, Zhuo Chen, Joshua T. Robinson, Hailiang Wang, Bo Zhang and Hongjie Dai*

Prof. H. Dai, G. Hong, S. M. Tabakman, J. T. Robinson, H. Wang, B. Zhang
Department of Chemistry, Stanford University, Stanford, California 94305
E-mail: hdai@stanford.edu
Dr. K. Welsher
Department of Chemistry, Princeton University, Princeton, New Jersey, 08544
Dr. Z. Chen
State Key Laboratory of Chemo/Biosensing and Chemometrics, College of Biology, Hunan University, Changsha, Hunan, 410082 (China)

Biological imaging in the near infrared (NIR) window enjoys advantages of deep photon penetration due to relatively high transparency, reduced scattering and low autofluorescence of biological tissues in the 0.8-1.4 μm range.[1-5] Organic dyes[6-9], quantum dots[10-12] and single-walled carbon nanotubes (SWNTs)[4] have been employed for *in vitro* and *in vivo* biological imaging in the NIR region. SWNTs are a group of one-dimensional (1-D) macromolecular fluorophores, with intrinsic bandgap fluorescence emission between 0.9-1.4 μm upon excitation in the visible or NIR.[4, 13-14] The large Stokes shift makes SWNTs ideal probes for biological imaging with high contrast and low background.[3] Thus far, SWNTs have been used as *in vitro* fluorescence tags for cell imaging,[15-16] *ex vivo* imaging of tissues and organs,[17-18] and *in vivo* imaging of normal organs as well as tumors.[19-20]

A common caveat of NIR fluorophores is the relatively low quantum yields compared to their counterparts (including organic dyes and quantum dots) with shorter emission wavelengths in the visible, which limits their imaging capabilities. For example, the IR800 dye (with a peak emission wavelength of 800 nm) exhibits a ~10% quantum yield,[21] and the indocyanine green (ICG) dye exhibits only ~4.3% quantum yield at the emission wavelength of 805 nm.[22] In contrast, molecules fluorescing at shorter wavelengths typically exhibit much higher quantum yields (IR700 ~24% at 700 nm emission[21]; cyanine-5 ~30% at 660 nm emission[23]; fluorescein ~91% at 521 nm emission[24]). SWNTs exhibit quantum yield ranging from 0.1% to 3%,[13, 25-26] due to intrinsic low-energy excitons that are optically forbidden,[27] and extrinsic quenchers such as metallic SWNTs in bundles[25] and oxygen in acidic environment[28-29]. To fully utilize the spectral advantages of NIR fluorophores, it is desirable to develop a general approach to enhancing the photoluminescence (PL) in the NIR, thus enhancing the biological imaging capability in this important spectral region.

Recently, we reported metal-enhanced fluorescence (MEF) by up to 14-fold for surfactant-coated SWNTs placed on top of nanostructured gold films synthesized purely in the solution phase (called 'Au/Au films').[30] Here, we demonstrated this Au/Au substrate as a powerful and general platform for NIR fluorescence enhanced (NIR-FE) cellular imaging using both SWNT and organic fluorescent labels. We used SWNTs functionalized by arginine–glycine–aspartic acid (RGD) to selectively tag U87-MG brain cancer cells over MCF-7 breast cancer cells, plated the cells on the Au/Au substrate, and observed a ~9-fold increase in SWNT fluorescence on U87-MG cells. This enabled high quality NIR molecular imaging of molecularly targeted cells using much shorter exposure times (~300 ms) than previously possible with nanotube fluorophores. With NIR-FE imaging, we were able to push the detectable limit of SWNT staining of cells down to an ultralow concentration of ~50 pM. Further, we observed different degrees of fluorescence enhancement for endocytosed, intracellular SWNTs vs. nanotubes on the cell membrane at the cell/gold interface, suggesting the possibility of observing transmembrane endocytosis of live cells based on the degree of fluorescence enhancement.

Also important is that our NIR-FE imaging of biological system is general for commonly used low quantum yield organic dyes including IR800. To our knowledge, this is the first fluorescence enhanced imaging of cells on Au nanostructures in the NIR. Previously, Ag substrates were used for fluorescence enhanced biological imaging in the visible with organic dyes.[31-33]

Cell-type selective staining and subsequent imaging of cells were carried out with RGD and IR800 conjugated SWNTs, water-solubilized by 25% DSPE-PEG(5k)-NH$_2$ (1,2-distearoyl-*sn*-glycero-3-phosphoethanolamine-N-[amino(polyethyleneglycol)

5,000]) and 75% C18-PMH-mPEG(90k) (poly(maleic anhydride-*alt*-1-octadecene)-methoxy(polyethyleneglycol)90,000). The RGD peptide ligand was linked to the amine groups on SWNTs for selectively binding to $α_vβ_3$-integrin positive U87-MG cells[34] over the $α_vβ_3$-integrin negative MCF-7 cells. Meanwhile, we also covalently attached IR800 dye molecules onto SWNTs to afford SWNT-IR800-RGD conjugates (Figure 1a). Upon excitation at 658 nm, the SWNT-IR800-RGD conjugate emitted in the range of 1000-1400 nm, due to the intrinsic bandgap photoluminescence of SWNTs (Figure 1e and 1f). Upon excitation at 785 nm, the conjugate emitted fluorescence in the 800-1000 nm range due to the attached IR800 molecules (emission tail also seen in Figure 1e and 1f). This unique SWNT-IR800 conjugate allowed for tagging of cells using two NIR fluorophores in two different imaging windows in the 800-1400nm range. Atomic force microscopy (AFM) imaging (Figure 1b) showed the SWNT conjugates with lengths ranging from 100 nm to 3 μm and an average length of ~1.0 μm.

We synthesized Au/Au films on quartz (Figure 1c) via solution phase growth, with optimal optical extinction in the NIR (Figure 1d) for the highest fluorescence enhancement of both SWNTs and IR800 placed on top of the gold film.[30, 35] For SWNT-IR800-RGD conjugates drop-dried from a solution onto both bare quartz and Au/Au film on quartz, photoluminescence versus excitation (PLE) spectra revealed fluorescence enhancement of both IR800 (only emission tail showing up at the upper left corner of Figure 1e and 1f in the spectral range) and SWNTs (all other peaks in the 1.0-1.4 μm emission window of Figure 1e and 1f) on the Au/Au film. The average enhancement of SWNT photoluminescence was ~10 times, and the enhancement was ~5 times for the IR800 dye attached to SWNTs. This result clearly showed the excellent capability of fluorescence enhancement by the Au/Au film for fluorophores emitting in the 0.8-1.4 μm NIR window including IR800 and SWNTs.

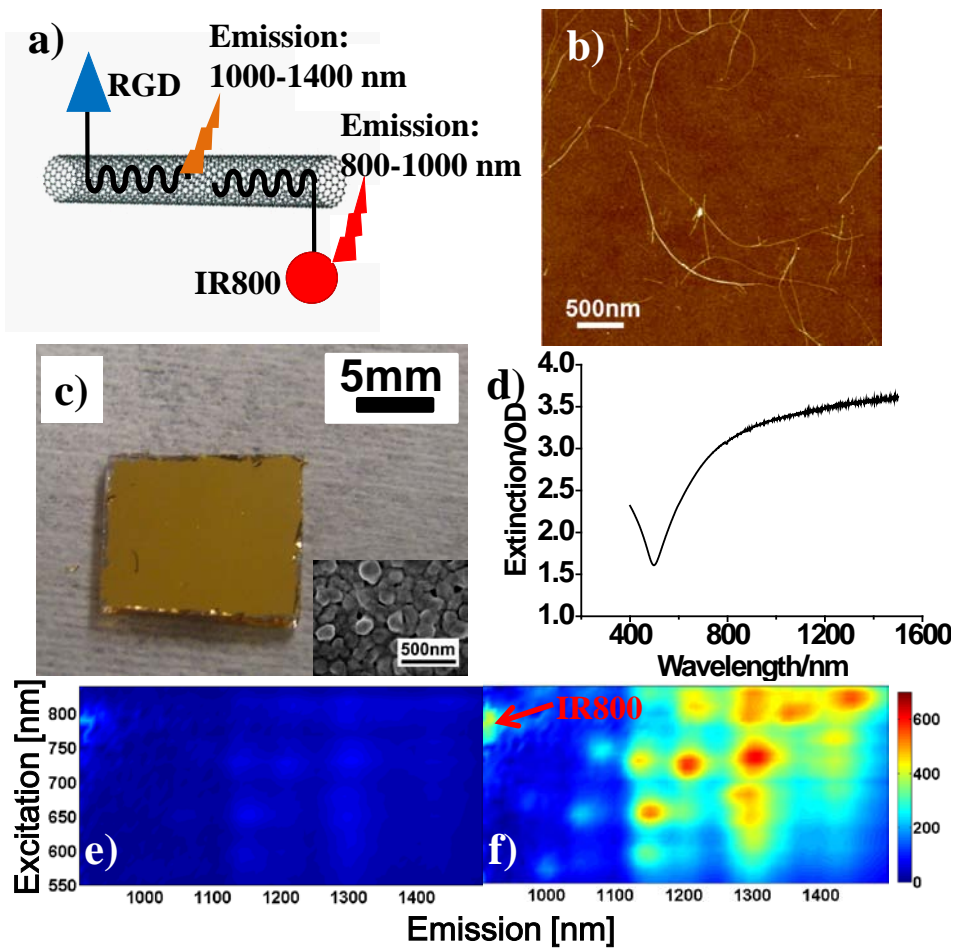

***Figure 1.*** a) A schematic drawing of the SWNT-IR800-RGD conjugate, with the emission ranges of SWNTs and IR800 dye labeled. b) An AFM image of SWNT-IR800-RGD conjugates deposited on a SiO$_2$ substrate. c) A digital photograph of a typical Au/Au substrate used for NIR-FE imaging of cells plated on this substrate. Inset showed an SEM image of the film. d) A UV-Vis-NIR extinction spectrum of the Au/Au film. e) A PLE spectrum of SWNT-IR800-RGD conjugates deposited on quartz. f) A PLE spectrum of SWNT-IR800-RGD conjugates deposited on top of an Au/Au film.

For targeted cell staining and imaging, U87-MG cells and MCF-7 cells were trypsinized and mixed with SWNT-IR800-RGD conjugates at a concentration of ~30 nM of SWNTs at 4 °C for 1h to prevent endocytosis during staining. The cells were split into two groups and placed onto a quartz microscope slide and Au/Au film respectively for immediate fluorescence imaging using an InGaAs camera. The $\alpha_v\beta_3$-integrin positive U87-MG cells treated with the SWNT-IR800-RGD conjugate showed ~9-fold higher SWNT fluorescence signal (green) on Au/Au (Figure 2a) than on quartz (Figure 2b), excited at 658 nm under a short exposure time of ~300 ms. Much longer exposure times (1~3s) were needed to obtain high quality SWNT-stained cell images on quartz, similar to previous biological imaging with SWNT fluorophores[19, 36]. The $\alpha_v\beta_3$-integrin negative MCF-7 cells on both Au/Au (Figure 2c) and quartz (Figure 2d) showed little SWNT fluorescence signal. The selectivity of RGD-SWNT labeling of cells, defined as the ratio of SWNT emission intensity of $\alpha_v\beta_3$-integrin positive U87-MG cells compared to that of $\alpha_v\beta_3$-integrin negative MCF-7 cells, was as high as ~17 for cells on Au/Au substrate (Figure 2g, higher than the positive/negative ratio of ~7 on quartz), suggesting highly selective staining and molecular imaging of cells with NIR-FE on the gold films. Note that the cells appeared round-shaped since they were imaged immediately after being placed on the substrates prior to adhesion. To show that these imaged cells were alive during and after imaging, we monitored the U87-MG cells in cell medium after increasing the temperature to 37 ºC in a temperature controlled imaging chamber with a 1 L/min $CO_2$ gas flow over a period of 6 hours after the first imaging. Both fluorescence image and optical image showed cells adhesion to the Au/Au surface at 37 ºC, suggesting live cells.

We also trypsinized and mixed cells with SWNT-IR800-RGD conjugates at 37 ºC (instead of 4 ºC as above) for 1h, a condition known to afford endocytosis of carbon nanotubes inside cells.[37-39] Single particle tracking of SWNTs in live cells has been studied by Strano and coworkers to reveal the mechanism of endocytosis at 37 ºC.[40-41] The U87-MG and MCF-7 cells thus treated were plated onto Au/Au and quartz for NIR imaging. In contrast to the ~9-fold enhancement observed for cells stained at 4 ºC, nanotube fluorescence in the $\alpha_v\beta_3$-integrin positive U87-MG cells treated at 37 ºC was enhanced by only ~2-fold on Au/Au film (Figure 2h) compared to on quartz substrate (Figure 2i). Also noticeable was the higher false-positive signal intensity in the $\alpha_v\beta_3$-integrin negative MCF-7 cells (Figure 2j, 2k), due to the expected increase of non-specific uptake of SWNTs by cells at 37 ºC than at 4 ºC.

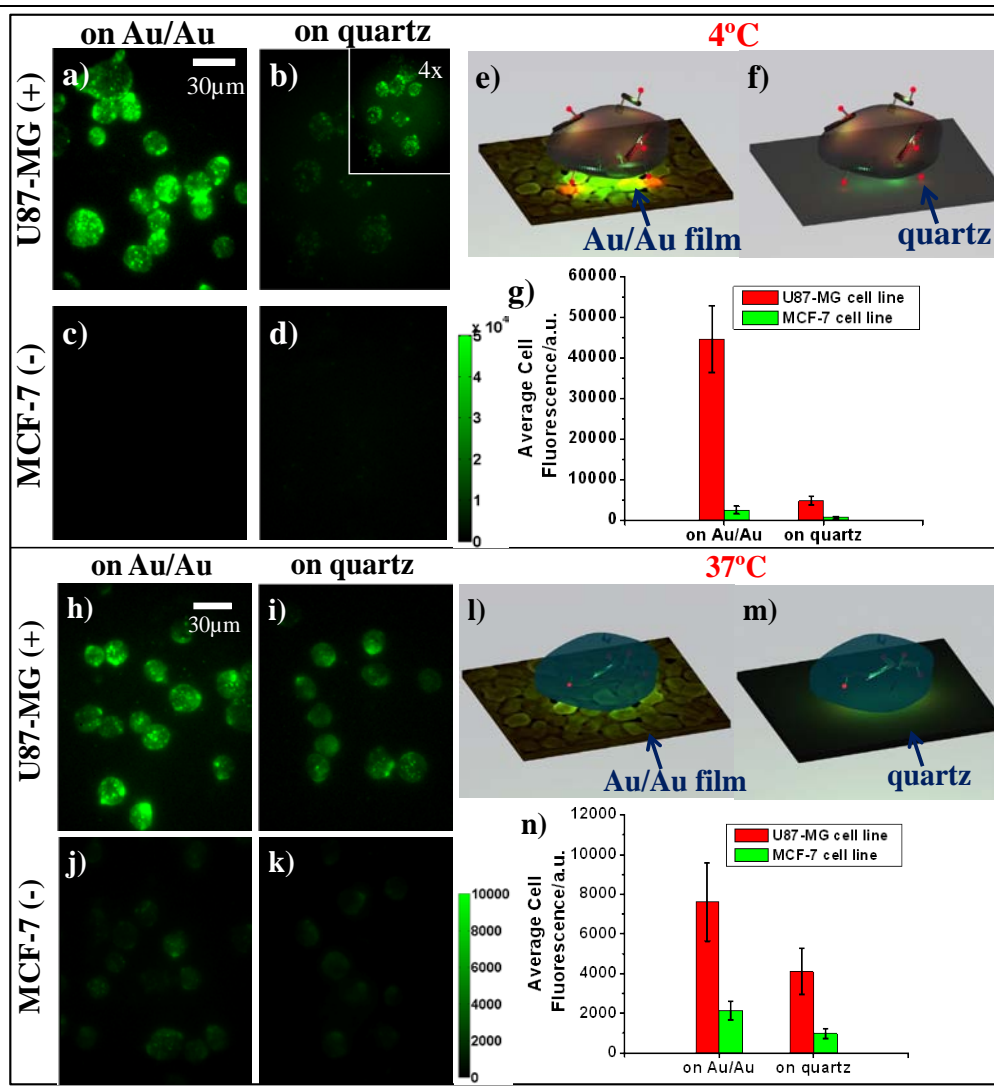

***Figure 2.*** NIR fluorescence images of a) SWNT-IR800-RGD stained U87-MG cells at 4ºC on Au/Au film, b) on quartz, and similarly treated MCF-7 cells c) on Au/Au and d) on quartz. The schematic drawings showed e) SWNTs sandwiched between cell membrane and gold were enhanced, compared to f) SWNT stained cell on quartz. g) A bar chart showing average cell fluorescence in each case of 2a)-d). NIR fluorescence images of h) SWNT-IR800-RGD stained U87-MG cells at 37ºC on Au/Au film, i) on quartz, and similarly treated MCF-7 cells j) on Au/Au and k) on quartz. The schematic drawings showed l) SWNTs endocytosed by the cell were hardly enhanced by gold, compared to m) SWNT stained cell on quartz due to increased distance between the fluorophores and the enhancing gold substrate. n) A bar chart showing average cell fluorescence in each case of 2h)-k). All false-colored

fluorescence images were taken under a 658 nm excitation with photons collected in the 1100-1700 nm region.

We attributed the fluorescence enhancement of the molecularly selective SWNT labels on U87-MG cells to coupling between the emissions of SWNT tags and surface plasmon modes in the Au/Au substrates. Resonance coupling between SWNT emission and re-radiating plasmonic modes in the Au/Au films shortened the radiative lifetimes of SWNTs, affording higher fluorescence quantum efficiency.[30, 42-43] It was found that surfactant-coated SWNTs closer to the Au/Au surface exhibited higher fluorescence enhancement, decaying when SWNTs were placed away from the surface with a half-decay distance of ~5 nm,[30] on the same order of cell membrane thickness.[44] At 4 ºC, most of the SWNTs were blocked from endocytotic uptake by the U87-MG cells, and SWNTs on the cell membrane interfacing with the Au/Au substrate were strongly coupled to the surface plasmonic modes in the gold film and thus responsible for the large, ~9-fold enhancement in fluorescence compared to on quartz (Figure 2e, 2f). On the other hand, when incubated at 37 ºC, SWNTs were endocytosed into the cytoplasms of the cells and hence spatially separated from the gold surface, giving a reduced fluorescence enhancement of ~2-fold (Figure 2l, 2m). Also interesting was that we monitored the U87-MG cells stained by SWNTs at 4°C in cell medium after increasing the temperature to 37 ºC in a 1 L/min $CO_2$ gas flow. We observed a ~ 6-fold decrease of SWNT fluorescence intensity in the U87-MG cells over time, from the fluorescence intensity taken right after 4 ºC staining. This decrease was due to transmembrane endocytosis of SWNTs at 37 ºC, which reduced the NIR-FE effect as nanotubes were further away from the Au/Au surface. This also confirmed that the cells were fully alive and functioning.

The SWNT-Au distance dependent fluorescence enhancement could also explain the measured increase in cellular targeting selectivity with cells on Au/Au vs. quartz substrate (Figure 2g). For integrin negative MCF-7 cells, the fluorescence signals detected were due to autofluorescence inside the cells and non-specific uptake effects, which were distributed through the cells in three dimensions. These non-specific signals were barely enhanced by the Au/Au substrate, while the specific SWNT signals on the integrin-positive U87-MG cells at the cell-gold interface were enhanced to the maximum degree due to proximity to Au. Thus, preferential enhancement of specific cell membrane surface fluorescence afforded more sensitive and selective imaging of cell membrane receptors. This effect was consistent with little enhancement in the cell labeling selectivity observed on Au/Au substrate for cells treated by SWNT-RGD at 37 ºC (Figure 2n). Interestingly, these results suggested that the distance dependent fluorescence enhancement effect could be used for tracking transmembrane behavior in live cells, since the thickness of cell membrane was on the same order of magnitude as the enhancement decay distance (~5 nm).

Besides the distance dependence of fluorescence enhancement, another reason for the observed increase in targeting selectivity on Au/Au film could be the nonlinearity of the enhancement effect. It was known that surface enhanced Raman scattering was non-linear to concentration, i.e., higher concentrations of analytes were usually enhanced more due to a better chance to occupy the enhancing 'hot spots'.[45-46] Therefore, the negative MCF-7 cells were not enhanced as much due to much fewer SWNTs on the membranes than the positive U87-MG cells, and as a result, targeting selectivity was magnified by this nonlinear effect.

Our NIR-FE imaging of cells was general for various NIR fluorescent organic dyes. We chose IR800 as a representative organic dye due to its wide use for biological imaging.[47-48] The IR800 dye molecules (shown as red circle in Figure 1a) bound to SWNTs deposited on the same Au/Au film exhibited a fluorescence enhancement by ~5-fold (Figure 1e vs 1f). We used SWNT-IR800-RGD conjugates to target $α_vβ_3$-integrin positive U87-MG cells and performed cell fluorescence imaging in the IR800 fluorescence channel. Comparing Figure 3a and 3c corresponding to SWNT-IR800-RGD stained U87-MG and MCF-7 cell lines respectively, one sees that the $α_vβ_3$-integrin positive U87-MG cells showed positive IR800 fluorescence signal (red) in the 790-820 nm region upon 785 nm excitation, while the $α_vβ_3$-integrin negative MCF-7 cells showed little IR800 signal under the same imaging condition. This again confirmed high specificity of molecular imaging and the coexistence of IR800 and RGD on SWNTs. Comparison of Figure 3a and 3b revealed a significant fluorescence enhancement on Au/Au vs. quartz by ~6-fold for IR800 labels on cells, with a positive/negative selectivity ratio of ~16 on Au/Au film vs. ~4 on quartz (Figure 3e). These results demonstrated the generality of NIR-FE imaging of cells for high molecular sensitivity and selectivity.

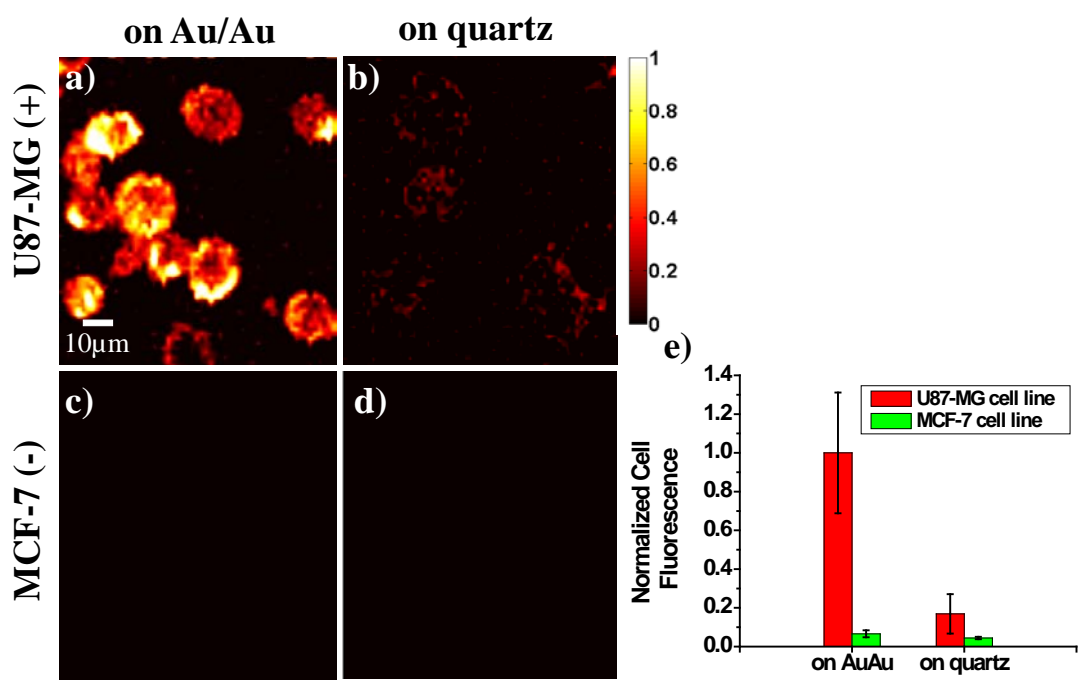

*Figure 3.* IR800 fluorescence images of SWNT-IR800-RGD stained U87-MG cells on a) Au/Au film and b) on quartz, and similarly treated MCF-7 cells on c) Au/Au and d) on quartz. Incubation was 1 h for both cell lines at 4 ºC. False-colored images were taken under 785 nm excitation with photons collected in the 790-820 nm region. The bar chart in e) showed normalized cell fluorescence in each case of 3a)-d).

Compared to organic NIR fluorophores, surfactant-solubilized SWNTs exhibit even lower quantum yield ranging from 0.1% to 3%.[13, 25-26] Relatively high concentrations of SWNTs (~60 nM) are typically needed for biological imaging, where the integrin binding sites to RGD on cell membranes were almost saturated.[16] To investigate the detection limit of NIR-FE cell imaging on gold, we labeled $\alpha_v\beta_3$-integrin positive U87-MG cells using different staining concentrations of SWNTs from 30 nM to 48 pM and imaged the cells on both Au/Au film (Figure 4, top row) and quartz slide (Figure 4, bottom row). A low detection limit of 1.2 nM SWNT-RGD was reached for labeled cells on quartz substrate, below which cells (visible in optical images in insets of Figure 4h, 4j) were hardly discerned in the SWNT fluorescence images. In contrast, on Au/Au substrates, SWNT fluorescent cell images were observed for cells treated by SWNTs down to ~48 pM in concentration, suggesting a 25-fold improvement in the detection limit on Au/Au film (Figure 4k).

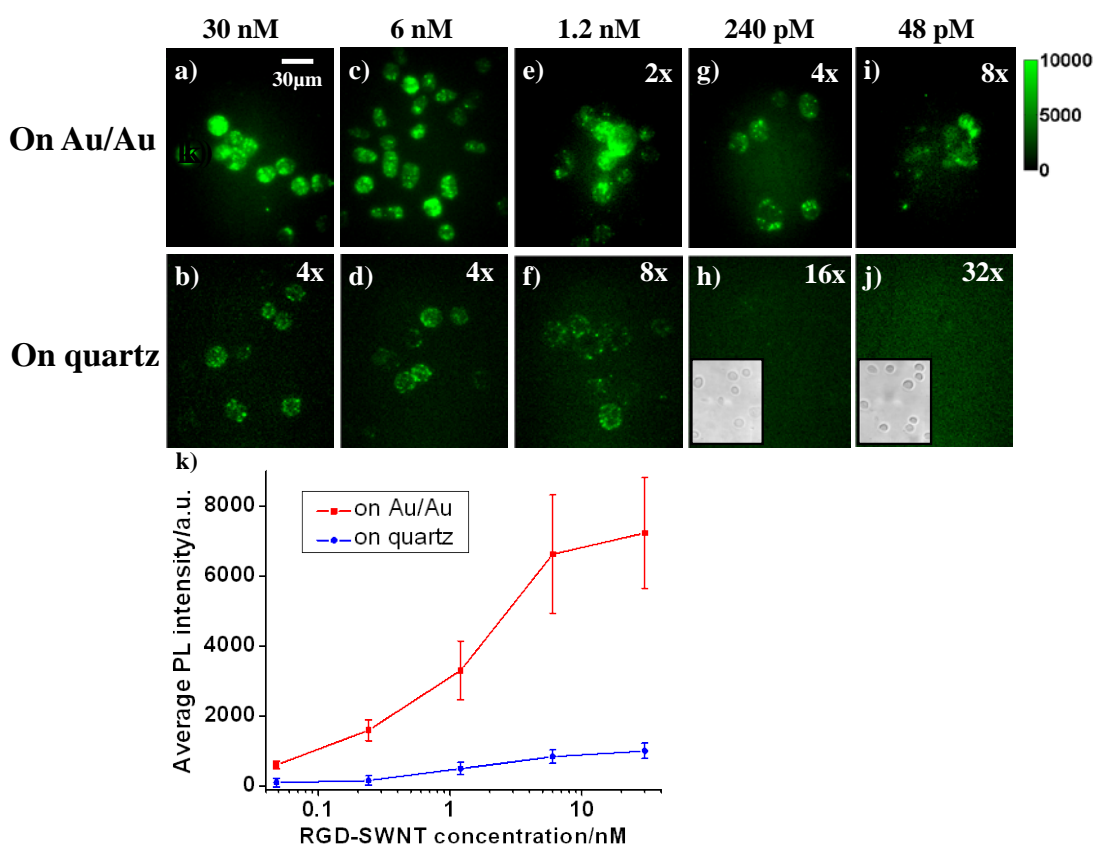

***Figure 4.*** SWNT NIR fluorescence images of $\alpha_v\beta_3$-integrin positive U87-MG cells, on Au/Au (top panel) and quartz (bottom panel), stained with different concentrations of SWNT-RGD conjugates: a)&b) 30 nM, c)&d) 6 nM, e)&f) 1.2 nM, g)&h) 240 pM and i)&j) 48 pM. Some of these images with weak fluorescence were autoscaled by a certain multiplier as labeled. Insets in h) and j) showed the corresponding optical images in the same field of view. All fluorescence images were false-colored. k) Cell titration curves showed the average fluorescence intensity of SWNTs in each stained U87-MG cell on Au/Au film (red curve) and on quartz (blue curve) at different staining concentrations.

In conclusion, we employed plasmonic gold substrates for the first time to perform near-IR fluorescence enhanced molecular imaging of cells in the 0.8-1.4 μm spectral window based on carbon nanotubes and organic fluorophores. The novel solution-grown gold substrate was general in enhancing both carbon nanotubes and infrared dye IR800, by ~9 times and ~6 times respectively, affording higher sensitivity and specificity of molecular cell imaging in the advantageous 0.8-1.4 μm spectral window. Cell labeling at different incubation temperatures blocked or allowed endocytosis of nanotube fluorophores, leading to observation of a distance dependent fluorescence enhancement inside cells. This effect could be used to observe transmembrane behavior of single NIR fluorophores in live cells when the fluorescence enhancement decay distance matches cell membrane thickness. Further possibilities with NIR-FE imaging include single molecule imaging and tracking of SWNTs or other NIR dyes on cell membrane, molecular imaging of low abundance cell membrane proteins, and even some *in vivo* NIR-FE imaging using smaller yet enhancing Au nanostructures as enhancing platform to serve in a fluidic system.